\documentclass[aps,prd,superscriptaddress,showkeys,showpacs,nofootinbib,twocolumn,a4paper]{revtex4-1}
\usepackage{ulem}
\usepackage{color}
\usepackage{graphicx}
\usepackage{appendix}
\usepackage{epsfig}
\usepackage[colorlinks=true,linkcolor= red, citecolor = cyan, urlcolor = teal ]{hyperref}
\usepackage{epstopdf}
\usepackage{amsmath}
\usepackage{subfig}
\usepackage{comment}
\usepackage{float}
\usepackage[dvipsnames]{xcolor}
\newcommand{\be}{\begin{equation}}
\newcommand{\ee}{\end{equation}}
\newcommand{\ba}{\begin{eqnarray}}
\newcommand{\ea}{\end{eqnarray}}
\newcommand{\nn}{\nonumber\\}

\begin{document}
		\title{Energy loss of heavy quarks in the presence of magnetic field}
	\author{Mohammad Yousuf Jamal}
	\email{mohammad@iitgoa.ac.in }
        \affiliation{School of Physical Sciences, Indian Institute of Technology Goa, Ponda-403401, Goa, India}	
	\author{Jai Prakash}
	\email{jai183212001@iitgoa.ac.in}
        \affiliation{School of Physical Sciences, Indian Institute of Technology Goa, Ponda-403401, Goa, India}

        \author{Indrani Nilima}
     \email{nilima.ism@gmail.com}   
        \affiliation{Department of Physics, Institute of Science, Banaras Hindu University (BHU), Varanasi, 221005, India}
        
        \author{Aritra Bandyopadhyay}
         \affiliation{Institut für Theoretische Physik, Universität Heidelberg, Philosophenweg 16, 69120 Heidelberg, Germany}

\begin{abstract}
{We study the heavy quark energy loss in the presence of a background magnetic field. The analysis considers the high magnetic field generated by spectators from initial hard collisions that were incorporated using the medium-modified Debye mass, determined from quark condensates at finite temperature and magnetic field via recent lattice quantum chromodynamics (LQCD) calculations. We analyse the impact of medium polarization on the heavy quark propagation in a quark-gluon plasma formed in relativistic heavy-ion colliders like RHIC and LHC. For simplification, we considered the static medium with constant temperature and magnetic field values. Then, we explore the nuclear modification factor ($R_{AA}$) at different magnitudes of magnetic field strengths at fixed temperatures.  The energy loss of heavy quarks significantly increases, leading to $R_{AA}$ suppression at higher magnetic field values.}\\
% We aim to study the energy loss of heavy quarks in the presence of the background magnetic field. 
%   To do so, we first investigate the effect of medium polarization on the propagation of charm and a bottom quark, considering an equilibrating quark-gluon plasma created in the relativistic heavy-ion colliders such as RHIC and LHC. The analysis is performed considering the high magnetic field produced due to the spectators from the initial hard collisions. Furthermore, we studied the nuclear modification factor ($R_{AA}$) for different values of magnetic field for the parameters relevant at RHIC and LHC energies. The energy loss of heavy quarks was found to increase significantly. The corresponding nuclear modification factor suppresses upto 15\% at the highest magnetic field strength studied ($eB$ = 0.6 $\text{GeV}^2$). \\
\end{abstract}
%%%%%%%%%%%%%%%%%%%%%%%%%%%%%%%%%%%%%%%%%%%%%%%%%%%%%%%%%%%%%%%%%%%%%%%
\maketitle
%%%%%%%%%%%%%%%%%%%%%%%%%%%%%%%%%%%%%%%%%%%%%%%%%%%%%%%%%%%%%%%%%%%%%%%
\section{Introduction}
The heavy quarks (HQs) produced in the heavy-ion collision (HICs) experiments play a crucial role in understanding the equilibrating phase, known as quark-gluon plasma (QGP), which is believed to be the phase of the early universe a microsecond after the big bang and presently, expected to exists in the core of neutron stars~ \cite{expt_rhic,expt_lhc, PHOBOS:2004zne, BRAHMS:2004adc, ALICE:2010khr, Fukushima:2020yzx}.
The HICs at the Large Hadron Collider (LHC) and Relativistic Heavy Ion Collider (RHIC) facilities provide an opportunity to create QGP in the laboratory. The main challenge in the study of the medium produced in HICs at various experimental facilities is its small size and short-lived nature. One has to extract information about the evolution of the QCD matter in HICs using the observables related to the final state particles detected, such as particle yields, their momentum, and angular distributions. In HICs, the HQs are one of the noble probes to study the transport properties. The HQs, while passing through the QGP, interact with the medium constitutes through various ways and lose energy.
The energy loss of HQs within the QGP medium is reflected in the suppression of their yields measured at high transverse momentum. Hence, it is of great interest for theoretical researchers to study the various mechanisms through which the HQs interact with the QGP medium.

There are several articles available where the authors have discussed the energy loss of HQs through different perspectives~\cite{Koike:1992xs, Baier:2000mf, Jacobs:2004qv, Armesto:2011ht, Majumder:2010qh,  Dokshitzer:2001zm, Jeon:2003gi,   Gyulassy:1999zd, Zakharov:2000iz,  Baier:2001yt, Mustafa:2003vh, DuttMazumder:2004xk, Meistrenko:2012ju, Burke:2013yra, Neufeld:2014yaa, Chakraborty:2006db, Adil:2006ei, Peigne:2005rk, Dusling:2009jn, Cho:2009ze,  Prakash:2024rdz, Han:2017nfz, Prakash:2023hfj,Prakash:2023wbs,Singh:2020fsj,Das:2022lqh}. 
In fact, a significant portion of the total energy loss of HQs in the equilibrated QGP is due to the interaction with the particles having momenta on the order of approximately the medium's temperature, $T$ denoted as ``hard modes", The energy loss of HQs resulting from hard interactions, originating from elastic collisions with the plasma constituents and inelastic collisions (radiative processes), has been extensively studied in various articles~\cite{Kurian:2020orp, Kumar:2021goi, Hong:2020diy, Saraswat:2015ena, Mustafa:2004dr, Qin:2007rn}. Additionally, within the plasma, there exist gauge fields known as ``soft modes" with momenta approximately on the order of $g_sT$, where ``$g_s$" denotes the strong coupling constant. The soft modes can be regarded as classical fields since they are highly occupied~\cite{Carrington:2016mhd}. When the HQ traverses through the medium, it interacts with both hard and soft modes of the QGP~\cite{Bandyopadhyay:2021zlm,Bandyopadhyay:2023hiv}. However, the soft contribution to the HQ's energy loss has received less attention in comparison to the contributions from the hard modes. This discrepancy arises from the fact that soft modes carry only a minor fraction of the total plasma energy. However, the interaction frequency of these classical fields with the HQ is non-negligible due to their high occupation number within the plasma. Physically, the soft component of energy loss corresponds to the interaction of the parton with soft collective excitations of the medium. Recent studies by the authors of Refs.~\cite{Carrington:2016mhd, Carrington:2015xca,Prakash:2023zeu} have demonstrated that the soft contribution to energy loss plays a vital role in the overall energy loss of the test parton within the medium and in the context of jet quenching phenomenology. The energy loss due to soft contributions can be studied through the medium polarization effect. When the HQ traverses through the medium, it induces the color field that provides a back-reaction against its motion, which can be taken into account by analyzing the polarization effects of the medium.

%Historically, this topic has been studied several times and is available in the literature. Earlier, the collisional energy loss suffered by the high energy partons due to the elastic scatterings off thermal quarks and gluons in QCD plasma studied by Bjorken ~\cite{bjorken1982energy}. Then the collisional energy loss formalism was developed by Thoma and Gyulassy~\cite{Thoma:1990fm} in which they studied it in terms of the longitudinal and transverse dielectric functions. A systematic framework was  constructed for the energy loss of both soft and hard momentum transfers within the finite temperature field theory approach in Refs. ~\cite{Braaten:1991jj, Mrowczynski:1991da, Thomas:1991ea}. Employing the ADS/CFT approach, the energy loss of moving HQs has been studied in Refs.~\cite{Fadafan:2008gb, Fadafan:2008uv, Fadafan:2012qu}. The anisotropic effects in the context of HQs energy loss have been studied in Refs. ~\cite{Romatschke:2004au, Baier:2008js, Carrington:2015xca} and considering the viscous corrections in Refs.~\cite{Jiang:2014oxa, Jiang:2016duz}. The radiative corrections to the energy loss of HQs are studied in Refs.~\cite{Mustafa:1997pm, Djordjevic:2003zk, Wicks:2007am,Abir:2011jb,Djordjevic:2008iz,Qin:2007rn, Cao:2013ita}. 

It has been further noticed that the study of HQ energy loss considering the effect of the magnetic field that is generated due to spectators in the initial HICs is not much studied and is mostly missing in the literature. That brings our attention to investigate it in the current article. Therefore, in this article, the aim is to study the polarization energy loss of the charm quark and bottom quark moving in the hot QGP medium in the presence of a background magnetic field. To do so, we follow the approach of the semi-classical transport theory, and hence, the basic structure of the analysis is mostly inspired by Refs.\cite{Thoma:1990fm, Koike:1992xs}. The incorporation of the magnetic field in the present work has been done through the medium-modified Debye mass that has been evaluated using the values of the quark condensates at finite temperature and magnetic field from a recent lattice quantum chromodynamic (LQCD) calculations~\cite{Bali1, Bali2}. The primary reason for using these LQCD-based quark condensates is to capture both the magnetic catalysis (MC) and inverse MC (IMC) effects in our estimations, which has been successfully carried out to explore the HQ potential within a magnetized medium in a very recent work~\cite{Nilima:2022tmz}. In HICs, as the plasma expands and cools down, the energy loss, which depends on the medium temperature, changes with time~\cite{Hong:2020diy}. Thus, we will use the medium-induced energy loss to investigate the energy loss effects on the $R_{AA}$. Hence, the effect of the magnetic field will be studied in the energy loss of HQs as well as in $R_{AA}$.

The manuscript is organized as follows. In Sec.~\eqref{sec:EL}, we provide the formalism for the energy loss of HQs in the presence of the background magnetic field. Section~\eqref{el:RaD}, is dedicated to the results and discussion. In Sec.~\eqref{el:SaF}, we summarize the present work and discuss future possibilities. Natural units are used throughout the text with $c=k_B=\hbar=1$. We use a bold typeface to indicate three vectors and a regular font to indicate four vectors. The center dot depicts the four-vector scalar product with the formula $g_{\mu\nu}={\text {diag}}(1,-1,-1,-1)$.

\section{Formalism}
\label{sec:EL}
The dynamics of HQs have intrigued researchers through various perspectives. Our focus lies in investigating their energy dissipation via medium polarization. Employing Wong equations, a framework describing classically colored charged particles interacting with the dynamic gluon fields, we delve into understanding the energy change in HQs. Wong equations \cite{Wong:1970fu}, defining the motion of a colored charged HQ in the dynamical gluon field $F_{a}^{\mu\nu}$, given as:

\ba
\frac{dX^{\mu}(\tau)}{d\tau} &=& V^{\mu}(\tau),\label{eq:1_1} \\
\frac{dQ^{\mu}(\tau)}{d\tau} &=&g \tilde{q}^{a}(\tau)F^{\mu\nu}_{a}(X(\tau)){V}_{\nu}(\tau),\label{eq:1_2} \\
\frac{d\tilde{q}^{a}(\tau)}{d\tau} &=& -gf^{abc}V_{\mu}(\tau)A^{\mu}_{b}(X(\tau))\tilde{q}_{c}(\tau),
\label{eq:wong}
\ea

in which $\tau$, $X^\mu (\tau)$, $Q^\mu (\tau)$, and $V^\mu (\tau)$ correspond to the proper time, position, momentum and velocity of the HQ, respectively, with a color charge $\tilde{q}_a$. Here, $A^\mu$ represents the gauge potential, $f^{abc}$ is the structure constant of the ${SU}(N_c)$ group with $a,b,c$ spanning $1,2,\dots, N_c^2-1$ for $N_c$ fundamental colors of quarks. The HQ passes through the medium induces the chromo-electric field that generates retarded forces. The polarization energy loss is then quantified by examining the work done by the HQ against these forces within the medium. Employing the formalism described in detail in Ref.~\cite{Carrington:2015xca}, Wong equations, along with the linearized Yang-Mills equation, provide the HQ energy loss as,
 \ba
    \frac{{\text{dE}}}{{\text {dt}}}&=&g_{s}\tilde{q}^a {{\bf v}}\cdot {{\it E}^{a,i}_{\text{ind}}}(t, {\bf x}={\bf v}t),
    \label{eq:el1st}
    \ea  
where $t = \gamma_L \tau $ (where $\gamma_L$ is a  Lorentz factor) and $ {\it E}^{a,i}_{\text{ind}}(t,{\bf x}={\bf v}t)$ is the induced chromo-electric field due to the motion of the HQ having energy, $\text{E}$  and velocity, ${\bf v}$. Now, ${\it E}^{a,i}_{\text{ind}}(K)$ in the Fourier space can be obtained as, 
\ba
{\it E}^{a,i}_{\text{ind}}(K) = i\omega \Delta^{ij}(K)j^{a,j}_{\text{ext}}(K),
\ea
where $K\equiv K^\mu=({\omega,{\bf k}})$. The gluon propagator, $\Delta^{ij}(K)$ takes the following form~\cite{YousufJamal:2019pen},
\ba
    \label{eq:pro}
    \Delta^{ij}(K) = [k^2\delta^{ij} - k^{i} k^{j} -\omega^2\epsilon^{ij}(K)]^{-1},
    \ea 
and the external current, $j^{a,j}_{\text{ext}}(K)$ is given by:
\ba 
j^{a,j}_{\text{ext}}(K)=\frac{i g_s \tilde{q}^a v^j}{\omega -{\bf k}\cdot{\bf v}+i 0^+}.
\ea
The dielectric permittivity, $\epsilon^{ij}(K)$ characterizing the medium can be expanded into transverse, $\epsilon_T(K)$ and longitudinal, $\epsilon_L(K)$ components, facilitating a deeper comprehension of induced fields. This decomposition unfolds as:
\ba
\epsilon^{ij}(K) = A^{ij}~\epsilon_T(K) + B^{ij}~\epsilon_L(K),
\label{eq:eplt1}
\ea
where $A^{\text{ij}}= \delta^{ij}-\frac{k^ik^j}{k^2}$ and $B^{\text{ij}}=\frac{k^ik^j}{k^2}$ are the usual transverse and longitudinal projections, respectively.
By further analyzing the induced fields in conjunction with the dielectric permittivity, the  energy change, $\frac{{\text{dE}}}{{\text {dx}}} (=\frac{1}{|{\bf v}|}\frac{{\text{dE}}}{{\text {dt}}})$ of a HQ due to medium polarization is obtained as,
\begin{align}
     \frac{{\text{dE}}}{{\text {dx}}}&=-\frac{\alpha_s C_F}{2 \pi ^2 |{\bf v}|}
    \int d^3{\bf k}\frac{\omega }{|{\bf k}|^2}\bigg\{\text{Im}\Big({\epsilon_{L}(K)}\Big)^{-1}\nonumber\\ 
    &+\left(|{\bf k}|^2 { |{\bf v}|}^2-\omega ^2\right)\text{Im}\Big({\omega ^2 \epsilon_T(K)-|{\bf k}|^2}\Big)^{-1}\bigg\}_{\omega = {\bf k}\cdot{\bf v}},\nn
    \label{eq:de1}
\end{align} 
where $C_F$ is the Casimir invariant of $SU(N_c)$. Now, Eq.\eqref{eq:de1} delves into various aspects depending on elements like the QCD coupling strength $\alpha_s$, the forms of $\epsilon_{L}(K)$ and $\epsilon_{T}(K)$. We shall discuss it in detail in the next sections.

\begin{figure}[ht]
		\centering
		\includegraphics[height=6.5cm,width=7.60cm]{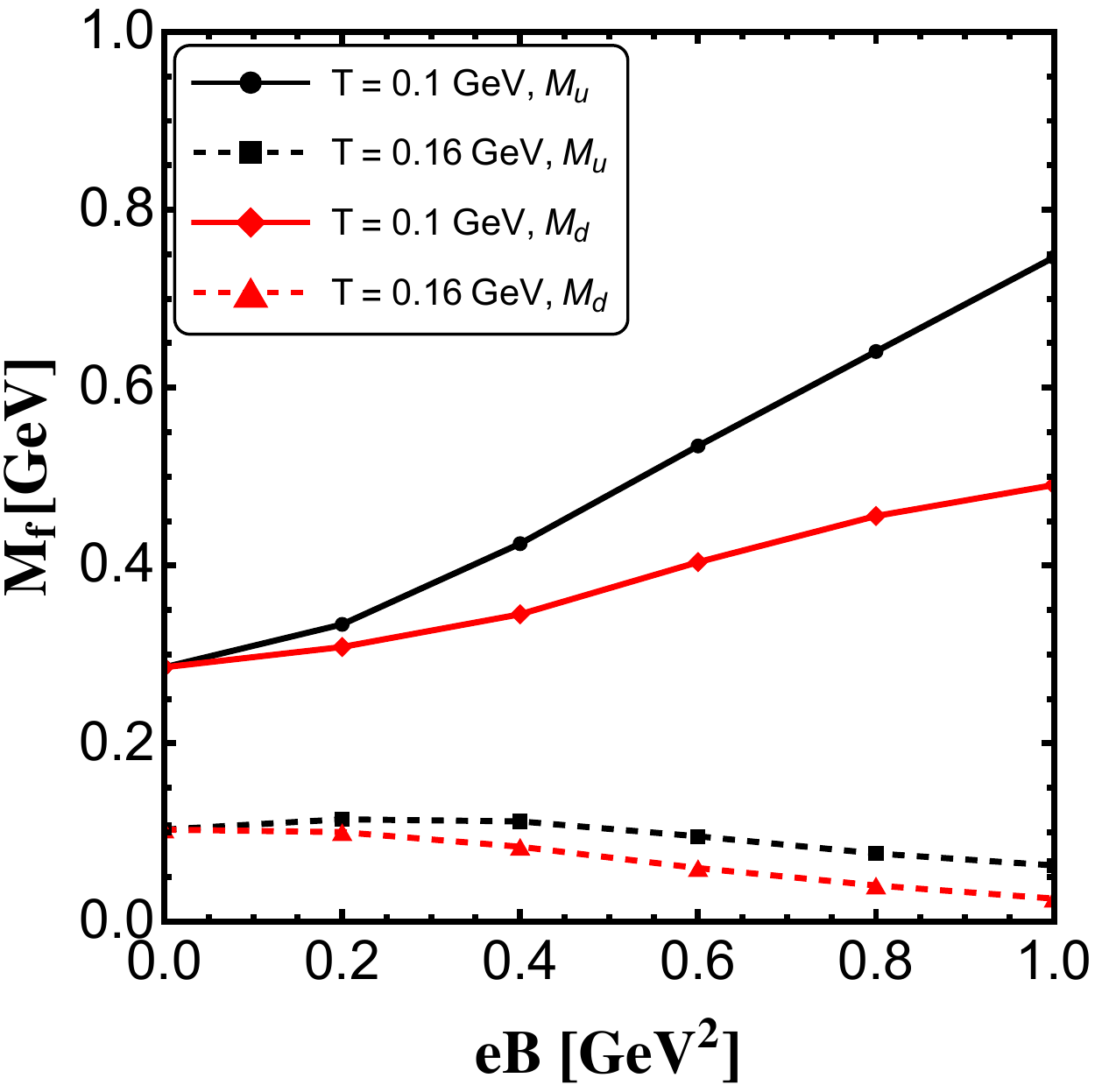}
		\caption{Variation of the LQCD-based effective quark mass (for up and down quarks) with the magnetic field at two different values of the temperature.}
		\label{fig:MvseB}
	\end{figure}

\subsection{Dielectric permittivity}

The dielectric permittivity has been computed previously by one of the authors in Refs.~\cite{Jamal:2017dqs, Kumar:2017bja}. Here, we are presenting the important steps for completeness. The $\epsilon^{ij}(K)$ of the QGP medium can be obtained from the gluon self-energy, $\Pi^{ij}(K)$ as,
   \ba
   \epsilon^{ij}(K)=\delta^{ij}-\frac{1}{\omega^2}\Pi ^{\text{ij}}(K).
   \label{eq:ep}
   \ea
   
 Further, $\Pi^{ij}(K)$ can be derived from the current induced due to the motion of the HQ~\cite{Mrowczynski:1993qm,Jamal:2022ztl},
   \ba
   \Pi^{ij}(K)=\frac{\delta j^{i}_{a, {\text ind}}(K)}{\delta A_{j, a}(K)}.
   \label{eq:p}	 
   \ea
 The $ j^{i}_{a, {\text ind}}(X)$ in the coordinate space is given as,
\ba
	j^{i}_{a, {\text ind}}(X)&=&g\int\frac{d^{3}p}{(2\pi)^3} u^{i}\{2N_c \delta f^{g}_a(p,X)+N_{f}[\delta f^{q}_a(p,X) \nn
	&-&\delta f^{\bar{q}}_a(p,X)]\},\label{indcurrent}
	\ea	
 where the change in the distribution functions of the medium particle, $\delta f^{i}$ is obtained by solving the linearized Boltzmann-Vlasov transport equation ~\cite{Romatschke:2003ms, Schenke:2006xu} written separately for each colour
channel as,
\ba
u^{\mu}\partial_{\mu} \delta f^{i}_a(p,X) + g \theta_{i} u_{\mu}F^{\mu\nu}_a(X)\partial_{\nu}^{(p)}f^{i}(\mathbf{p})=0,
\label{transportequation}
\ea
where the index $i$, refers to the particle species (quark, anti-quark, and gluon) and $u^{\mu}=(1,\mathbf{u}=\mathbf{p}/|\mathbf{p}|)$, is the four-velocity of the plasma particle. $\theta_{i}\in\{\theta_g,\theta_q,\theta_{\bar{q}}\}$ having the values $\theta_{g}=\theta_{q}=1$ and $\theta_{\bar{q}}=-1$. Next, solving Eq.\eqref{transportequation} for $\delta f^{i}_a(p,X)$ and using it Eq.\eqref{indcurrent} in the Fourier space we obtain $\delta j^{i}_{a, {\text ind}}(K)$. Further employing it in Eq.\eqref{eq:p} we get,
	\ba
	\Pi ^{ij}(K)&=& m_D^2\int \frac{d\Omega }{4 \pi }u^{i} u^{l}\Big\{\delta ^{lj}+\frac{u^{j} k^{l}}{\omega -{\bf k}\cdot {\bf u}}
	\Big\},
	\label{eq:pi}
	\ea
where $m_D$ is the Debye screening mass. Now, using Eq.\eqref{eq:eplt1}, \eqref{eq:ep} and \eqref{eq:pi} with proper contractions, one can get $\epsilon _L(K)$ and $\epsilon _L(K)$ as~\cite{ Jamal:2020fxo, Jamal:2021btg, Jamal:2020emj},
    \ba
    \epsilon _L(K)=1+m_D^2\left[\frac{1}{k^2}-\frac{\omega}{2k^3} \ln \left(\frac{\omega+k }{\omega-k}\right)\right],
    \label{eq:eL}
    \ea 
    \\
    and
    \ba
    \epsilon _T(K)&=&1-\frac{m_D^2}{2 \omega k} \bigg[\frac{\omega}{k}+\Big(\frac{1}{2}-\frac{\omega^2}{2k^2}\Big)\ln \left(\frac{\omega+k }{\omega-k}\right)\bigg].\nn
    \label{eq:eT}
    \ea	
{ At this point, it is important to highlight that in our analysis, we have made the assumption, that the effect of the magnetic field is solely encapsulated in $m_D$. This subsequently means that in Eq.~\eqref{indcurrent}, the background state or the phase space is considered to be isotropic. This approximation is driven by our specific goal of examining the MC/IMC effects coming through lattice on the HQs energy loss, which can be comprehensively captured in $m_D$. We recognize the possibility of achieving greater generality through the inclusion of structural anisotropies within the correlation functions. However, it is beyond the scope of the current analysis and is designated for an upcoming future investigation.} Next, we shall discuss the Debye screening mass and the strong coupling in the presence of the magnetic field.

 % \begin{figure*}[ht]
	% 	\centering
	% 	\includegraphics[height=5.5cm,width=6.60cm]{charm_rhic.pdf}
 %  \hspace{3mm}
	% 	\includegraphics[height=5.5cm,width=6.60cm]{charm_LHC.pdf}
	% 	\caption{Energy loss for charm quarks is plotted as a function of momentum (p)  at T = 360 MeV (right panel) and at T = 480 MeV (right panel) at different values of the magnetic field. }
	% 	\label{fig:EL_C}
	% \end{figure*}

 % \begin{figure*}[ht]
	% 	\centering
	% 	\includegraphics[height=5.5cm,width=6.60cm]{bottom_rhic.pdf}
 %  \hspace{3mm}
	% 	\includegraphics[height=5.5cm,width=6.60cm]{bottom_LHC.pdf}
	% 	\caption{Energy loss for bottom quark is plotted as a function of momentum (p) at T = 360 MeV (right panel) and at T = 480 MeV (right panel) at different values of the magnetic field. }
	% 	\label{fig:EL_B}
	% \end{figure*}

\subsection{Screening mass and strong coupling in the presence of magnetic field}
\label{sec:mD}
The screening mass depicted in Eqs.\eqref{eq:pi}, \eqref{eq:eL} and \eqref{eq:eT} is given by:
\ba
m^{2}_D &=& 4\pi \alpha_s \bigg(-2N_c \int \frac{d^3  p}{(2\pi)^3} \partial_p f_g(p)\nn &-& N_f \int \frac{d^3 p}{(2\pi)^3} \partial_p \left(f_q(p)+f_{\bar q}(p)\right)\bigg).
\ea
%The effects of the magnetic field in our analysis are encapsulated through $m_D$, and we now delve into a detailed discussion of its composition. 
The computation of $m_D$ in a magnetized medium has been investigated recently in Refs.~\cite{Bandyopadhyay:2016fyd, Bonati:2017uvz, Singh:2017nfa, Kurian:2017yxj, Kurain:2019, Karmakar:2019tdp, SG_VC}. In this investigation, we evaluate $m_D$ using semi-classical transport theory~\cite{Carrington, Agotiya:2016bqr, Jamal:2018mog}, specifically tailored for a magnetized medium~\cite{Nilima:2022tmz}. The expression for $m_D$ in a magnetized medium, within the framework of semi-classical transport theory, is articulated as follows:

\ba
m_{D}^{2}(T,eB) &=& g_s^2(T,eB) T^2 \frac{N_c}{3} + \dfrac{g_s^2(T,eB) |q_feB|}{\pi^2T} \nn &\times& \int_{0}^{\infty} dp_z \sum_{l=0}^{\infty}(2-\delta_{l0})f^l(1-f^l),
\ea
{where $l$ represents the Landau levels, summation of which is originating from the quantization of the transverse part of the phase space integral $d^3p$ in presence of an external magnetic field, i.e. $\int\limits_0^\infty\int\limits_0^\infty dp_xdp_y \rightarrow |q_feB|\sum_{l=0}^{\infty}(2-\delta_{l0})$.} $g_s(T,eB)$ is the temperature and magnetic field dependent coupling constant considered in the present study.~\footnote{Unlike ref.~\cite{Nilima:2022tmz} where only temperature dependence was considered.} The temperature and magnetic field dependence is incorporated in $g_s(T,eB)$~\cite{Ayala:2015bgv,Ayala:2018wux} as per the expression:
\ba
\begin{aligned}
g_s^2(T,eB) &= 4\pi\alpha_{s}(T,eB) \\
&= \frac{4\pi\alpha_s(T)}{1+\alpha_s(T)\frac{11N_c-2 N_{f}}{12\pi}\ln \left(\frac{4\pi^2 T^2}{4\pi^2T^2+eB}\right)},
\end{aligned}
\ea
where
\ba
\alpha_{s}(T) = \frac{6 \pi}{\left(11N_c-2 N_{f}\right)\ln \left(\frac{2\pi T}{\Lambda_{\overline{\rm MS}}}\right)},
\ea
and $\Lambda_{\overline{\rm MS}}$ represents the $\overline{\rm MS}$ renormalization scale. This kind of temperature and magnetic field-dependent coupling constant captures the behaviors of the competing scales $T$ and $eB$ and increases with the enhancement of the dominant scale. The Fermi-Dirac (FD) distribution functions for quarks, denoted as $f^l$, are defined as {$f^l = \frac{1}{e^{\beta E^l_f}+ 1}$}, where $E^l_f=\sqrt{p_{z}^{2}+m^{2}_f+2l |q_f eB|}$ {captures the Landau quantized dispersion relation with $l$ representing the Landau levels}. The fractional charge of $u$ and $d$ quarks, $q_f$, is $+\frac{2}{3}$ and $-\frac{1}{3}$ respectively. Finally, from here onward we will use $m_{D}\equiv m_{D}(T,eB)$.

Now, to enrich our expression for the Debye screening mass in a magnetized medium, we go one step further than most of the earlier studies of the Debye screening mass in a magnetized medium~\cite{Bandyopadhyay:2016fyd, Bonati:2017uvz, Singh:2017nfa, Kurian:2017yxj, Kurain:2019, Karmakar:2019tdp, SG_VC} and incorporate both the effects of Magnetic Catalysis (MC) and Inverse Magnetic Catalysis (IMC) in our system. It is a well-known fact that the value of the quark condensates ($\langle q\bar{q}\rangle_f$) increase with increasing magnetic field far away from the chiral transition temperature, $T_c$, which has been termed as MC and on the contrary, the same values of $\langle q\bar{q}\rangle_f$ starts to decrease with an increasing magnetic field near $T_c$ showing signs of IMC. Recently, in Ref.~\cite{Nilima:2022tmz}, these two effects have been captured through the medium dependent constituent quark mass $M_f(T, B)$, the medium dependence of which comes from the LQCD predicted values for the normalized quark condensates $\langle q \bar{q}\rangle_f (T,eB)$~\cite{Bali1, Bali2}. The condensates $\langle q \bar{q}\rangle_f (T,eB)$ are normalized in a way that brings down the constituent quark mass down from $M_f(T=0)$ to bare mass $m_f$ with increasing temperature and simultaneously varies from 1 to 0 during the chiral phase transition at $eB=0$. The relation between the medium-dependent constituent quark mass $M_f$ and the quark condensates $\langle q \bar{q}\rangle_f$ can subsequently be expressed as, 
\ba
M_f(T,eB)&=&M_f(T=0,eB=0)\times \langle q{\bar q}\rangle_f(T,eB) + m_f\nn
&\approx&M_f(T=0,eB=0)\times \langle q{\bar q}\rangle_f(T,eB)~,
\label{M_TB}
\ea
which in turn modifies our dispersion relation in a magnetized medium as, 
\begin{equation}
\bar{E}_f^l = \sqrt{p_z^2+2l|q_f eB|+M_f(T,eB)^2}.
\label{disq_imc}
\end{equation}

In Fig.~\ref{fig:MvseB} we show the variation of the constituent quark mass $M_{f=u,d}$ with magnetic field using LQCD based $\langle q{\bar q}\rangle_f(T,eB)$ from Ref.~\cite{Bali1} in Eq.~(\ref{M_TB}). One can clearly notice that at a lower temperature, {\it i.e.,} $T=100$ MeV, the constituent mass is enhanced with increasing $eB$, mapping the MC effect of the quark condensates. On the other hand, the constituent quark mass near $T_c$, {\it i.e.,} at $T=150$ MeV, first increases and subsequently decreases with increasing $eB$, a clear reflection of the IMC effect of the quark condensates. With these modifications, the final expression for the Debye screening mass comes out to be: % 
\begin{eqnarray}\label{6}
m_{D}^{2}&=&g_s^2(T,eB) T^2 \frac{N_c}{3} +\sum_f \dfrac{g_s^2(T,eB)|q_feB|}{\pi^2T}\nn
&\times&\int_{0}^{\infty} dp_z\sum_{l=0}^{\infty}(2-\delta_{l0})~{f}^l(\bar{E}_f^l)~\left(1-{f}^l(\bar{E}_f^l)\right).\nn
\end{eqnarray}
{So, the effects of the magnetic field have been taken into account within the Debye mass through i) the Landau quantized dispersion relation $\bar{E}_f^l$, ii) the Landau quantized phase space integral $|q_feB|\sum_{l=0}^{\infty}(2-\delta_{l0})$ and iii) the running coupling $g_s(T,eB)$. This magnetic field effects within $m_D$ then propagates in the energy loss (Eq. \eqref{eq:de1}) through the polarization functions $\epsilon_L$ and $\epsilon_T$.} Hence, solving Eq.~\eqref{eq:de1} using Eqs. \eqref{eq:eL}, \eqref{eq:eT}, and \eqref{6}, we obtain the energy loss of HQs due to medium polarization in the presence of the magnetic field.

 \begin{figure}[ht]
		\centering
		\includegraphics[height=6.5cm,width=7.60cm]{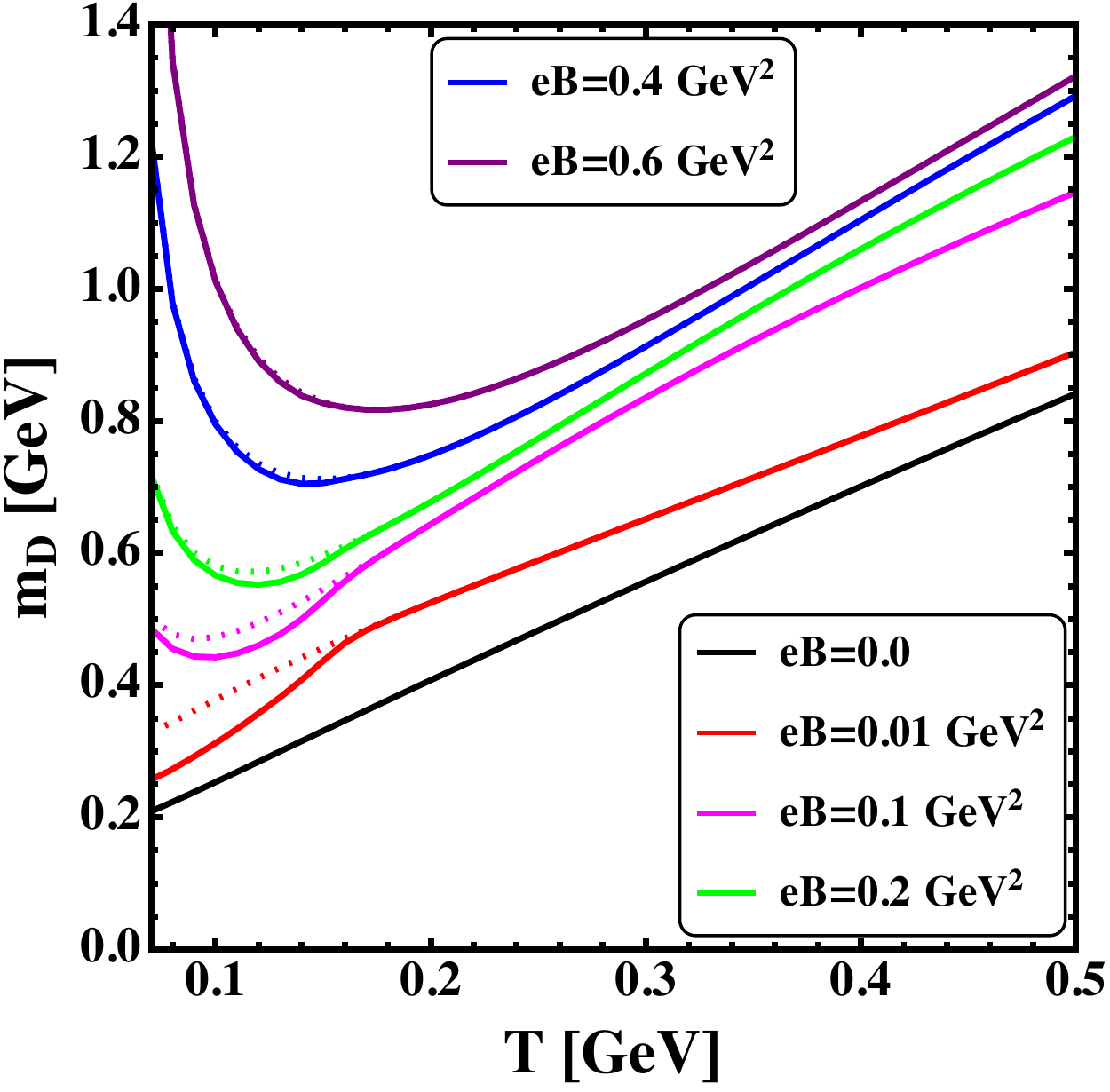}
		\caption{Variation of Debye screening mass ($m_D$) with the temperature at different values of the magnetic field ($eB$). The dotted lines represent the values of $m_D$ with a bare quark mass ($m_f$) for corresponding values of $eB$.}
		\label{fig:Md}
	\end{figure}

 \begin{figure*}[ht]
	\centering
		\includegraphics[height=6.6cm,width=7.60cm]{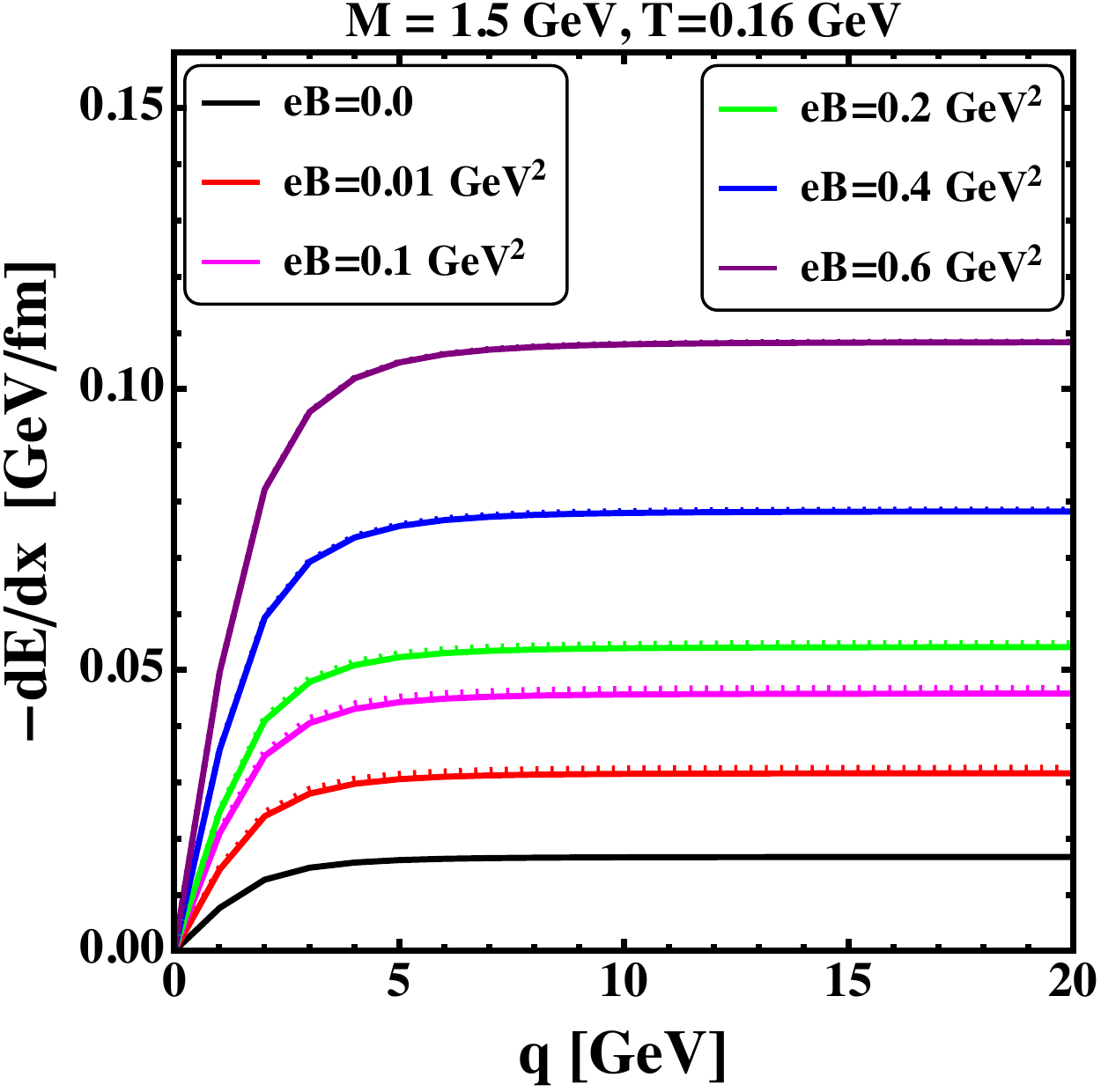}
  \hspace{5mm}
            \includegraphics[height=6.6cm,width=7.60cm]{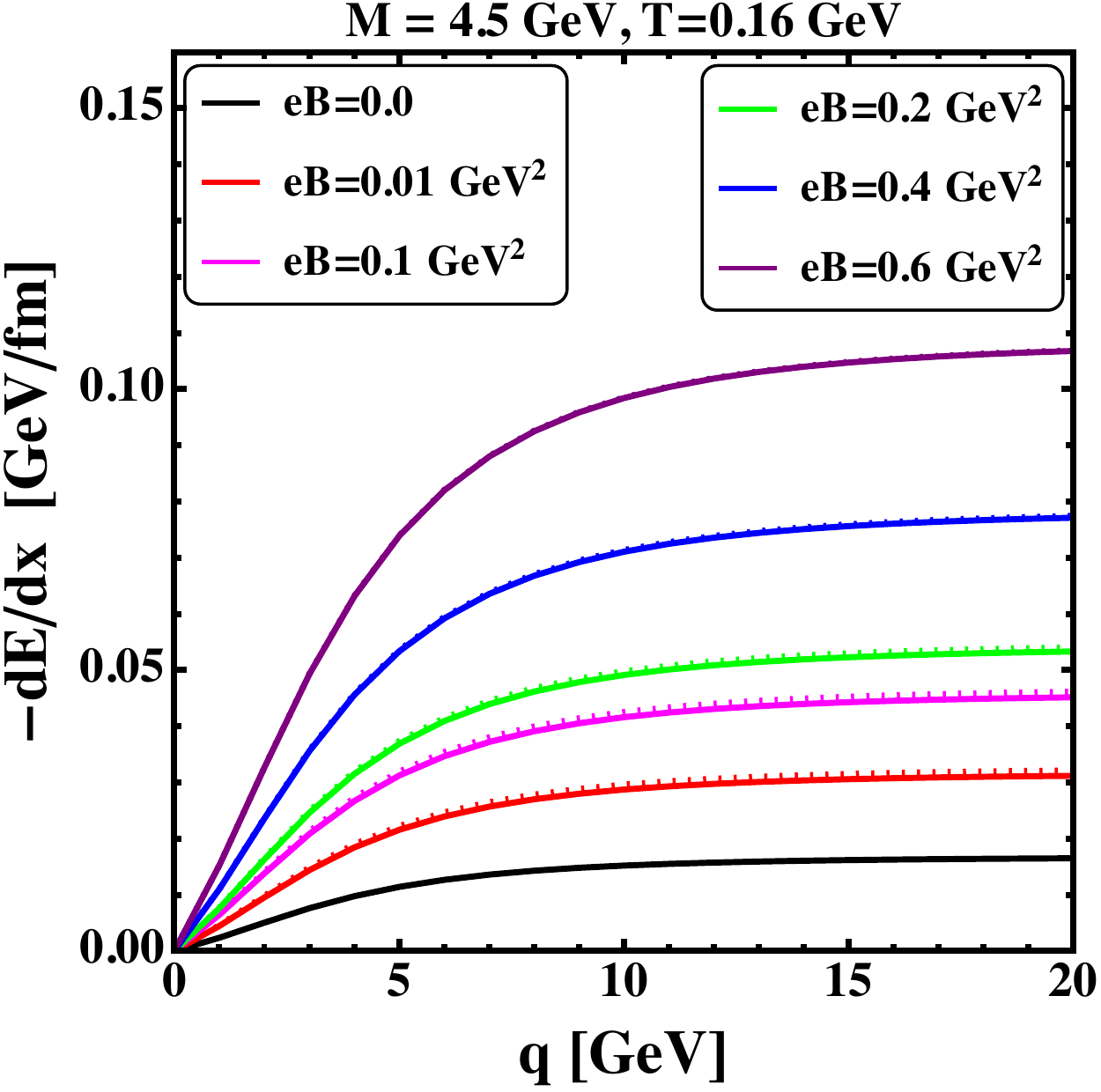}
	\caption{Variation of energy loss of charm (left panel) and bottom (right panel) with momenta at different values of the magnetic field ($eB$) at temperature, $T$ = 0.16 GeV. The dotted lines represent the corresponding values for $eB$ with a bare light quark mass ($m_f$).}
		\label{fig:EL_T16}
	\end{figure*}

 \begin{figure*}[ht]
	\centering
		\includegraphics[height=6.6cm,width=7.60cm]{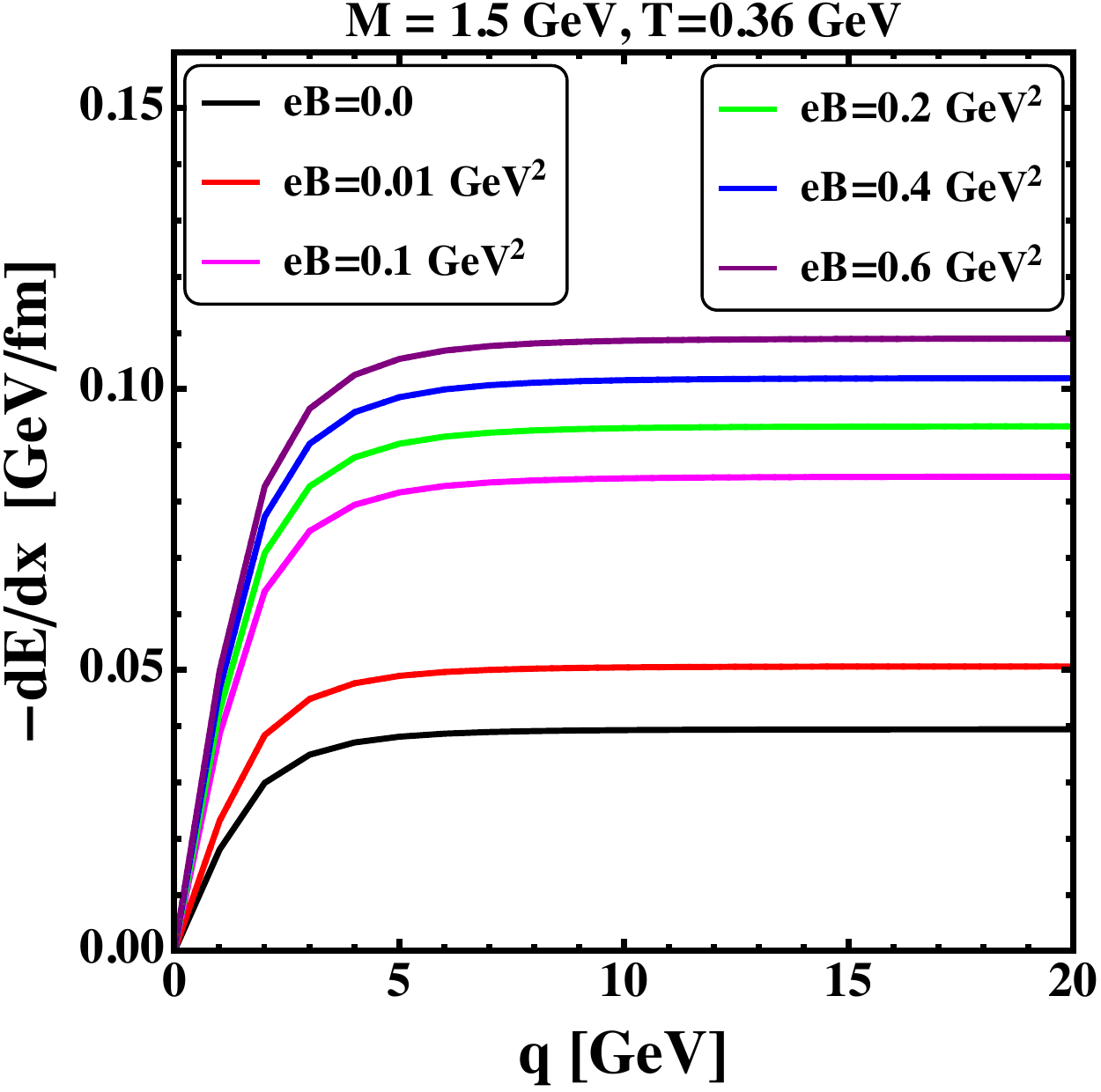}
  \hspace{5mm}
            \includegraphics[height=6.6cm,width=7.60cm]{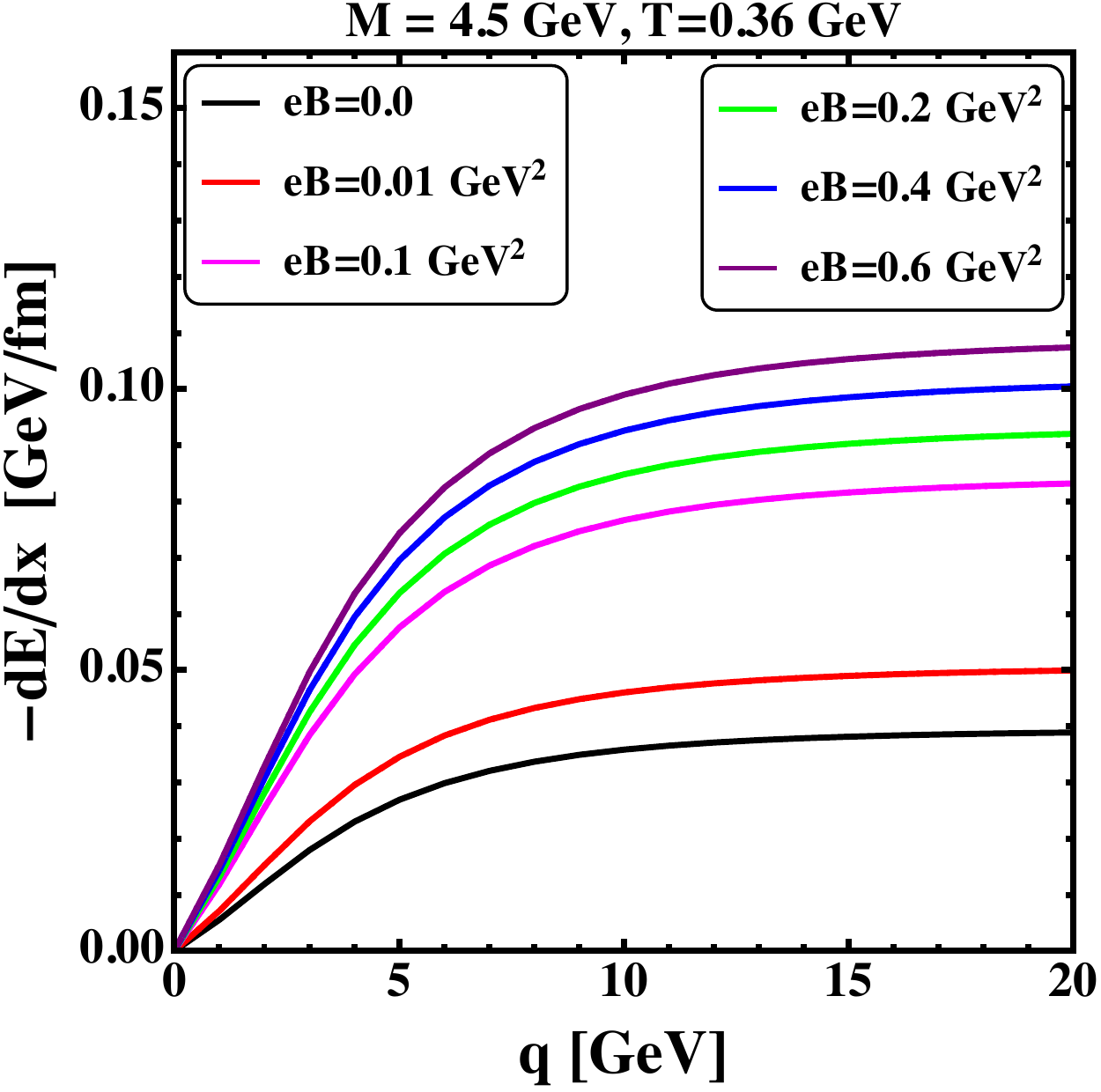}

            \caption{Variation of energy loss of charm (left panel) and bottom (right panel) with momenta at different values of the magnetic field ($eB$) at temperature, $T$ = 0.36 GeV.}
		\label{fig:EL_T36}
	\end{figure*}

 \begin{figure*}[ht]
	\centering
		\includegraphics[height=6.6cm,width=7.60cm]{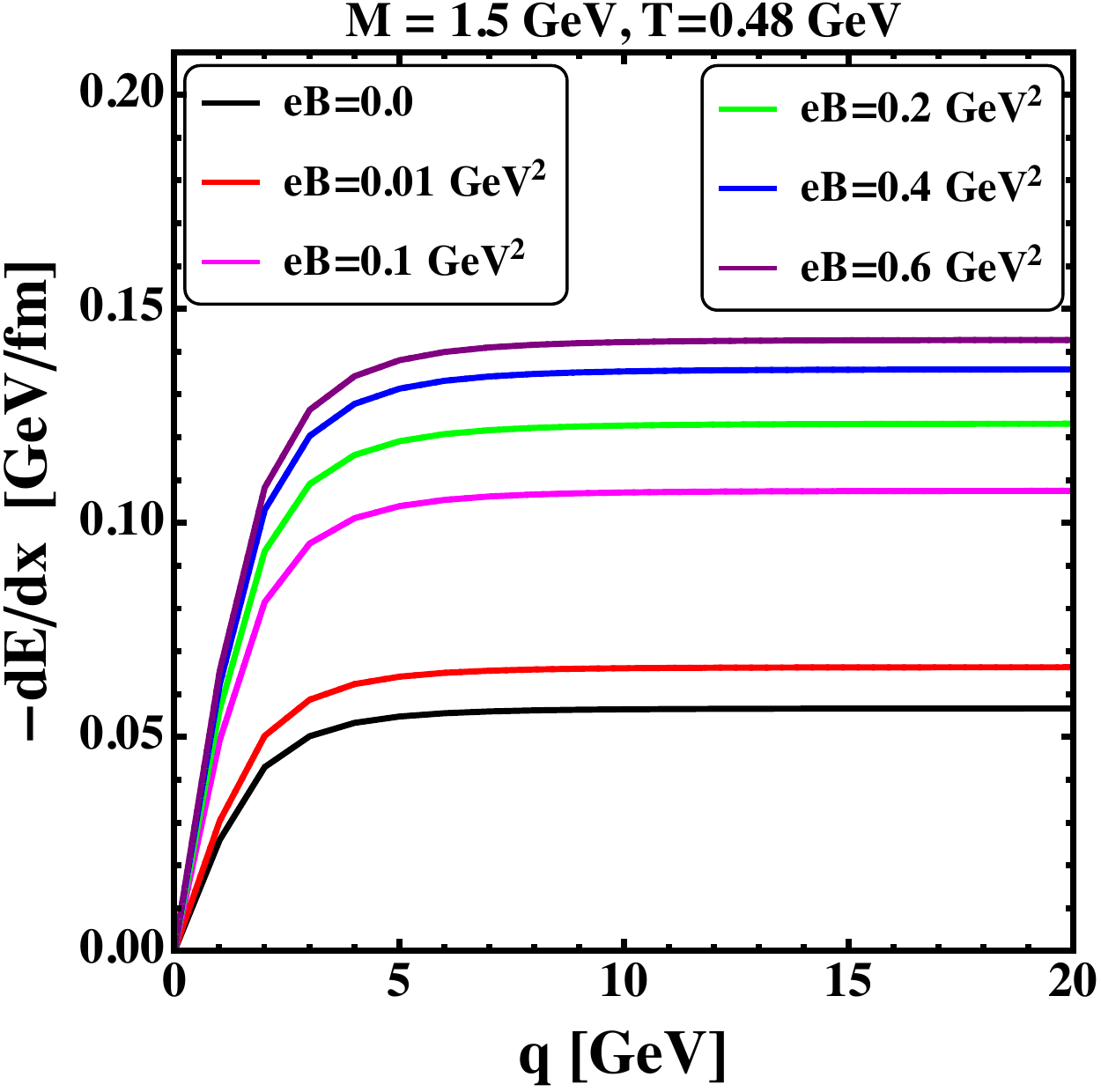}
  \hspace{5mm}
            \includegraphics[height=6.6cm,width=7.60cm]{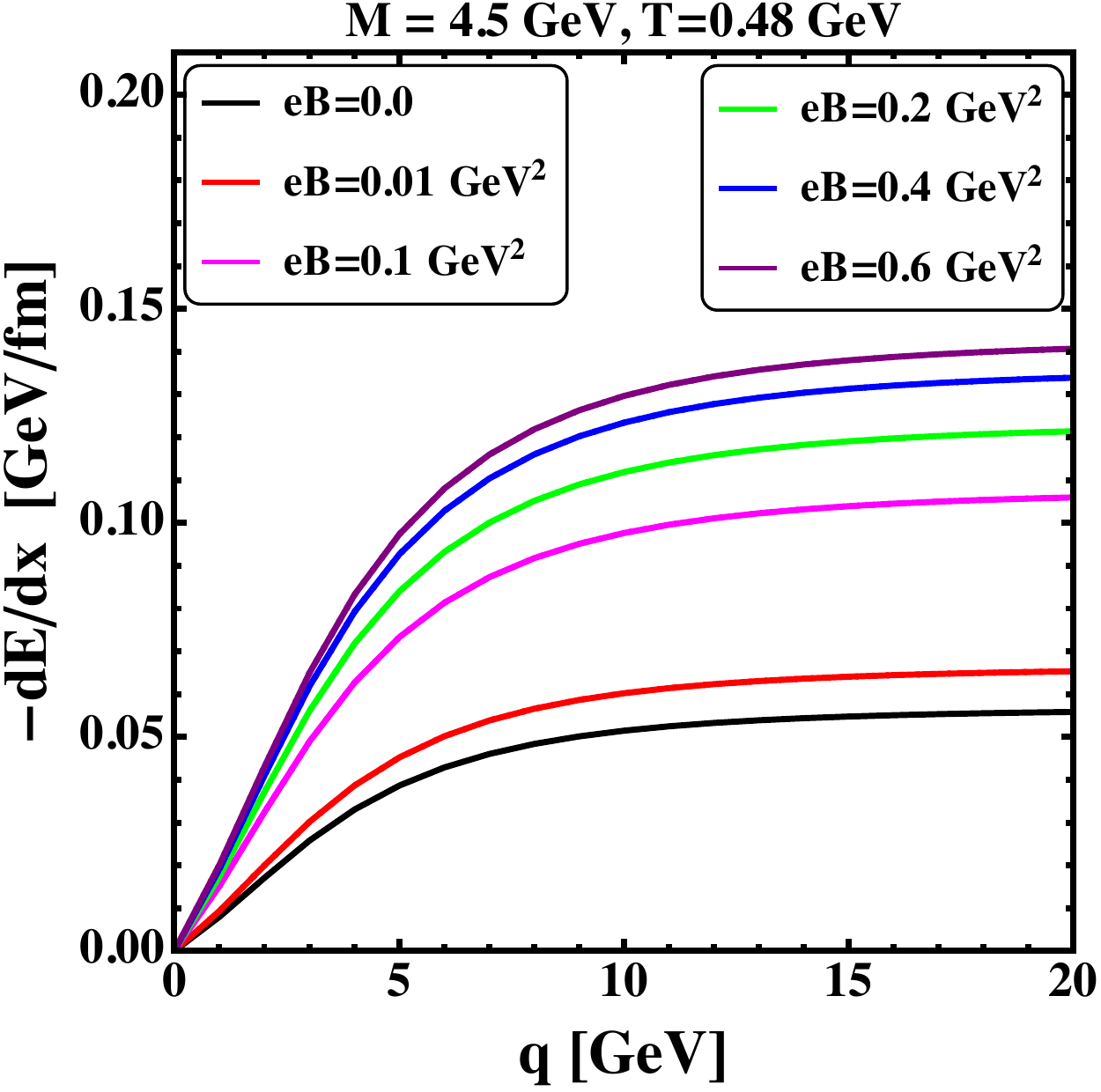}

            \caption{Variation of energy loss of charm (left panel) and bottom (right panel) with momenta at different values of the magnetic field ($eB$) at temperature, $T$ = 0.48 GeV.}
		\label{fig:EL_T48}
	\end{figure*}

\begin{figure*}[ht]
		\centering
        \includegraphics[height=5.6cm,width=5.60cm]{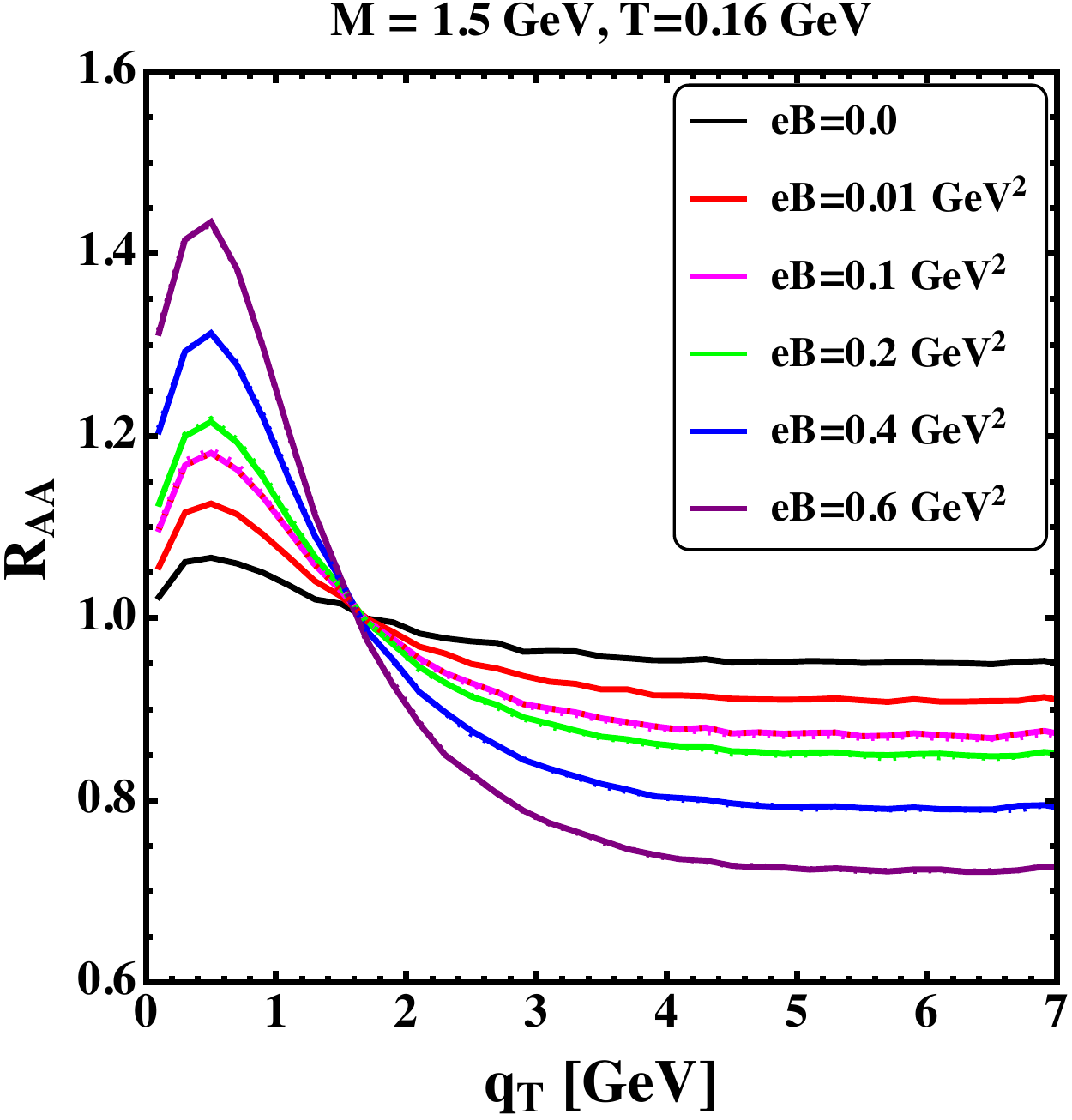}
        \hspace{-2mm}
	\includegraphics[height=5.6cm,width=5.60cm]{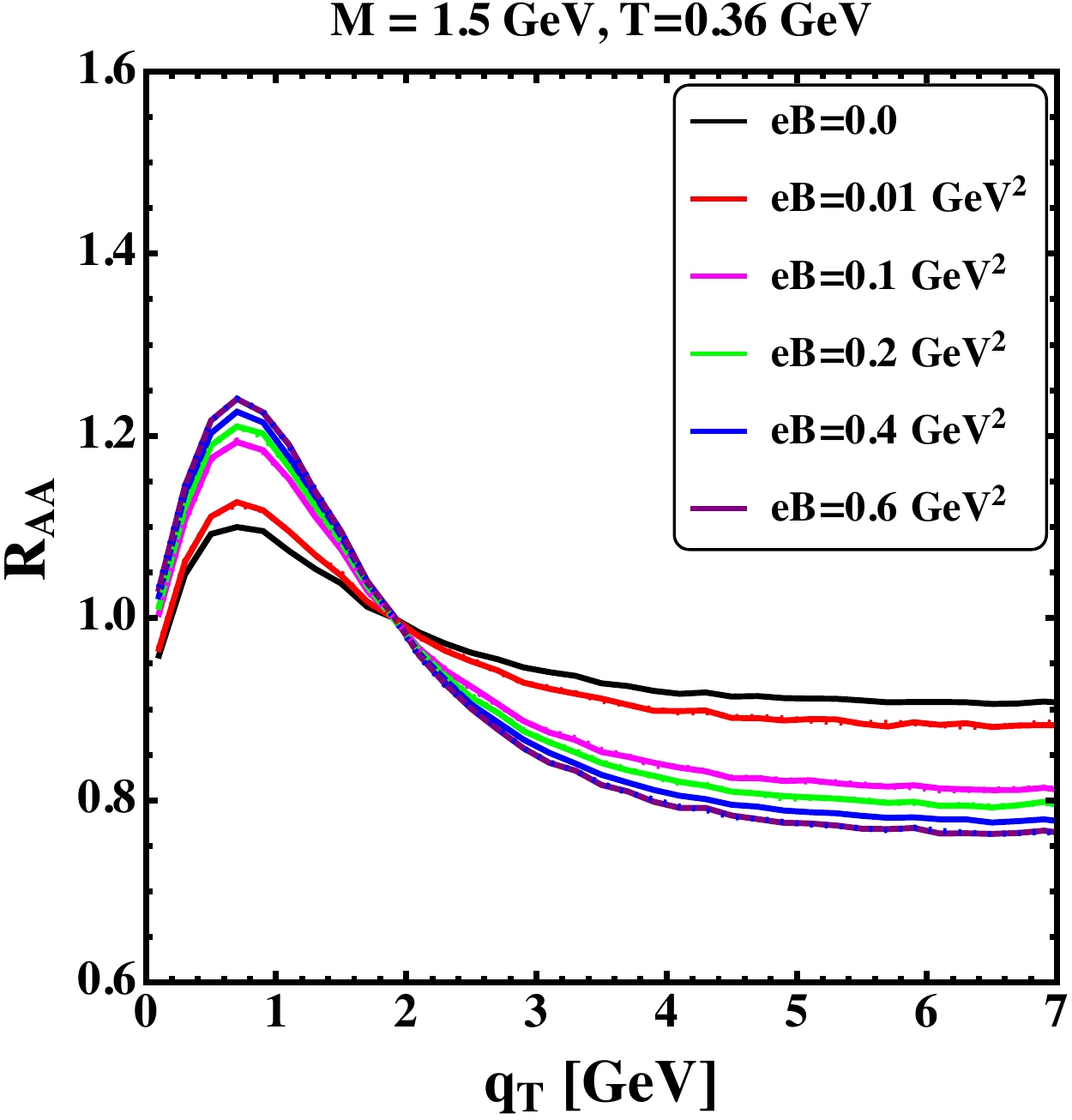}
 \hspace{-2mm}
	\includegraphics[height=5.6cm,width=5.60cm]{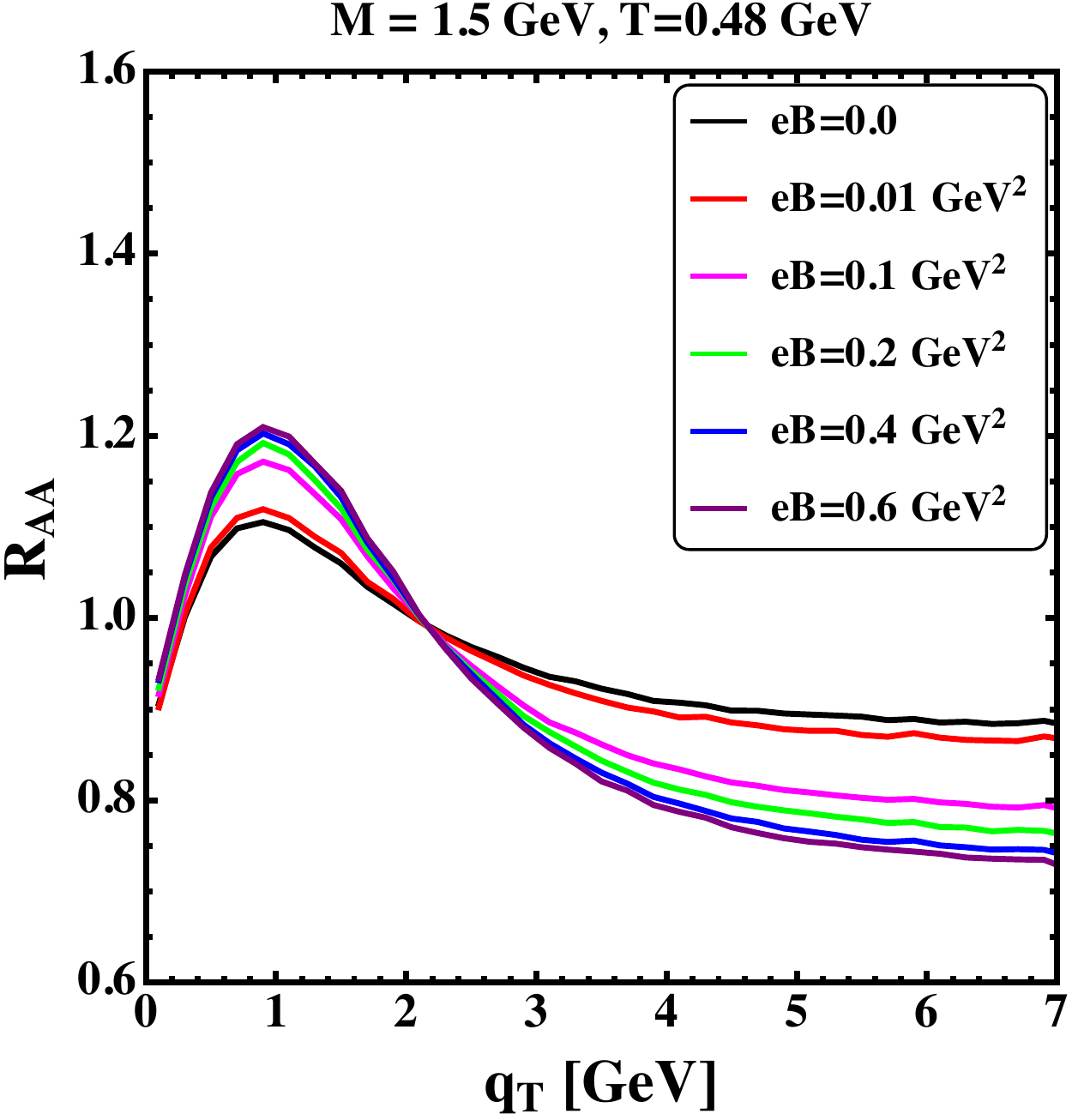}
 \caption{Variation of $R_{AA}$ of charm quark at T=0.16 GeV (left panel) and T=0.36 GeV (right panel) with transverse momenta at different values of the magnetic field ($eB$). The dotted lines represent the values of $m_D$ with a bare quark mass ($m_f$) for corresponding values of $eB$.}
		\label{fig:RAA_C}
	\end{figure*}

\section{Results and discussion}
\label{el:RaD}
In this section, we shall discuss our results regarding the HQ energy loss due to medium polarization in the presence of a magnetized medium incorporating {LQCD-based} quark condensate. In the present analysis we have considered, $N_c = 3$, $N_f = 2$ and $\Lambda_{\overline{\rm MS}} = 0.176$ GeV ~\cite{Haque:2014rua}. Throughout our results section we choose six different values of the external magnetic field, i.e. $eB=0$ (black curve), $eB=0.01$ GeV$^2$ (red curve), $eB=0.1$ GeV$^2$ (magenta curve), $eB=0.2$ GeV$^2$ (green curve), $eB=0.4$ GeV$^2$ (blue curve) and $eB=0.6$ GeV$^2$ (purple curve), capturing the whole spectrum of weak to strong magnetic field regime. For the HQ energy loss and $R_{AA}$ we show variations at three different temperatures, i.e. $T=0.16$ GeV (relevant to transition temperature $T_c$), $T=0.36$ GeV (relevant to RHIC)  and $T=0.48$ GeV (relevant to LHC). The energy loss of the charm and bottom quarks have been plotted against their momenta at different values of the magnetic field, as shown in various figures, which we analyze next.  

\subsection{Energy loss}
We start our analysis by showing the variation of the Debye screening mass, $m_D$, with respect to the temperature at different values of the magnetic field in Fig.~\ref{fig:Md}. The magnetic field enters the analysis through the particle distribution function and carries the MC and IMC profiles through the medium-dependent constituent quark mass and coupling constant that, in turn, appears in $m_D$. It is observed from Fig.~\ref{fig:Md} that both with the increase in the magnetic field values or in the temperature values, $m_D$ increases. The dotted lines for each value of the magnetic field represent the corresponding values of $m_D$ when we consider a bare quark mass instead of the medium-dependent constituent quark. One can clearly notice that the MC/IMC effect of the LQCD-based constituent mass affects the $m_D$ more within the lower temperature regime (close to $T_c$) and for lower values of the magnetic field.

 Figs.~\ref{fig:EL_T16}, \ref{fig:EL_T36} and \ref{fig:EL_T48} depict how the energy loss of charm (left panels) and bottom (right panels) quarks behave as a function of its momentum at different magnetic field strengths for three different temperatures: $T = 0.16$ GeV (Fig.~\ref{fig:EL_T16}), $T = 0.36$ GeV (Fig.~\ref{fig:EL_T36}) and $T = 0.48$ GeV (Fig.~\ref{fig:EL_T48}). The results indicate that the energy loss initially rises with momentum and then saturates. Furthermore, the magnetic field has shown a significant impact on the energy loss experienced by HQs in the QGP, as shown in the same figure. It is found that the HQ energy loss remarkably increases in the presence of a magnetic field for all three temperatures considered here. The effect is more pronounced at higher temperatures, as shown in Fig.~\ref{fig:EL_T48}. Specifically, the energy loss experiences an increase of approximately up to 180\% at $T = 480$ MeV (for the highest $eB$ value considered), and similar increments are observed in Figs.~\ref{fig:EL_T16} and \ref{fig:EL_T36}. These observations suggest that the magnetic field provides more hindrance to the motion of HQs in the QGP medium. Figs.~\ref{fig:EL_T16}, \ref{fig:EL_T36} and \ref{fig:EL_T48} also indicate that the energy loss of HQs increases as the strength of the magnetic field increases. The results for the bottom quark show a similar pattern to those of the charm quark. However, the energy loss of bottom quarks saturates at higher momentum due to their heavier mass. We also notice that the effect of MC/IMC through the LQCD-based medium-dependent constituent quark mass is negligible for the HQ energy loss. It was expected for higher temperature values, as the chiral symmetry gets restored in those temperatures. But even for a temperature close to $T_c$, {\it i.e.,} $T=160$ MeV, one can see that the difference between HQ energy loss with and without considering the MC/IMC effect is really small, as depicted by the dotted lines in Fig.~\ref{fig:EL_T16}. Hence, one can conclude that the MC/IMC effect which is well reflected in the Debye mass for lower values of $eB$ and close to $T_c$, eventually gets well suppressed in the HQ energy loss.
 
\subsection{Nuclear modification factor, $R_{AA}$}

The nuclear modification factor, denoted by $R_{AA}$, is a measure of the modification in the momentum spectra of HQs produced in heavy-ion collisions. It is defined as the ratio of the final momentum spectra of HQs, denoted by $f_{\tau_f}(\text{q}_T)$ at a given final time $\tau_f$ (where $\text{q}_T$ is the transverse momentum of the charm quark), to the initial momentum spectra of HQs, denoted by $f_{\tau_i}(\text{q}_T)$ at the initial time $\tau_i$. In mathematical notation, $R_{AA}$ is expressed as follows,

\begin{align}
 R_{AA}(\text{q}_T)=\frac{f_{\tau_f} (\text{q}_T )}{f_{\tau_i} (\text{q}_T)},
 \label{eq:raa}
\end{align}
 the value of the nuclear modification factor, $R_{AA}\rightarrow 1$ indicates that there is no significant interaction between the HQs and the surrounding medium. To calculate $R_{AA}$, the final momentum spectra of HQs, denoted by $f_{\tau_f}(\text{q}_T)$, are obtained after a time evolution of $\tau_f = 7$ fm/c, while the initial momentum spectra of HQs, denoted by $f_{\tau_i}(\text{q}_T)$.
{Initially, we presume that the momentum distribution of c-quarks is the prompt one obtained through Fixed Order + Next-to-Leading Log (FONLL) QCD, which describes the D-mesons spectra in pp collisions after fragmentation \cite{Ruggieri:2018rzi, Cacciari:2005rk, Cacciari:2012ny, Braaten:1994bz, Cacciari:2003zu}. This is written as follows:
\begin{equation}
\frac{dN}{d^2q_T} = \frac{x_0}{(x_1+q_T)^{x_2}},
\end{equation}
the parameters used in the calculation of $R_{AA}(q_T)$ for HQs are $x_0 = 6.365480 \times 10^8$, $x_1 = 9.0$ {GeV}, and $x_2 = 10.27890$. Normalization of the spectrum is important but not relevant to calculating the $R_{AA}(q_T)$, which is a ratio of the final over the initial spectrum given in Eq.\eqref{eq:raa}, and this is unaffected by the overall normalization since the number of heavy quarks is conserved during the evolution.} Next, the
 numerical calculations of $R_{AA}(q_T)$ for this process is performed using stochastic Langevin dynamics, a technique used to calculate the momentum evolution of HQs in a hot QCD medium. This method has been previously employed in the literature  ~\cite{Prakash:2021lwt, Das:2013kea, Ruggieri:2022kxv, Das:2017dsh, Plumari:2020eyx},
 \begin{align}
&dx_i=\frac{q_i}{E} dt,\nonumber \\
& \label{lang}dq_i=-\gamma q_i\, dt+C_{ij}\rho_j\sqrt{dt}.
\end{align}
 Here, $dq_i$ and $dx_i$ represent the shifts in momentum and position of the HQs in the medium, respectively, and $dt$ is the time step. The Langevin equation incorporates two types of forces acting on the HQs: the dissipative force and the stochastic force. The dissipative force is responsible for the drag effect and is denoted by $\gamma$, while the stochastic force contains the thermal noise and is represented by $\rho_j$, also known as white noise. The stochastic force exhibits properties such as $\langle \rho_i \rho_j \rangle = \delta_{ij}$ and $\langle \rho_i \rangle = 0$. $C_{ij}$ is the covariance matrix expressed as follows: \cite{vanHees:2005wb,vanHees:2007me, Scardina:2017ipo, Cao:2015hia}, 
\begin{align}
C_{ij} = \sqrt{2D_0}\left(\delta_{ij}-\frac{q_iq_j}{q^2}\right)+\sqrt{2D_1}\frac{q_iq_j}{q^2}.
\end{align}
In the analysis, the HQs are considered under the static limit, where $q \rightarrow 0$. In this limit, one can express $D_0 = D_1 = D$, where $D_0$ and $D_1$ represent the transverse and longitudinal diffusion coefficients, respectively. Doing so, $C_{ij}$ can be expressed as $\sqrt{2D} \delta_{ij}$, where $D$ represents the diffusion coefficient of the HQs in the medium. Now, the diffusion coefficient is calculated from drag coefficients using the fluctuation-dissipation theorem, $D=\gamma ET$ \cite{Walton:1999dy, Mazumder:2013oaa, Moore:2004tg}. Here $\gamma$ is calculated by using Eq. \eqref{eq:de1} as estimated in Refs. \cite{Prakash:2023zeu,Debnath:2023zet}, 
\ba
\gamma=-\frac{1}{q}\left(\frac{{\text {dE}}}{\text {dx}}\right).
\ea

 Next, we have plotted $R_{AA}$ for the charm quark as a function of transverse momentum, q$_T$ in Fig.~\ref{fig:RAA_C}  for $T$ = 0.16 GeV (left panel), $T$ = 0.36 GeV (middle panel) and $T$ = 0.46 GeV (right panel). In the presence of a magnetic field, the $R_{AA}$ is found to suppress at high q$_T$ (above 2 GeV)
The transition from $R_{AA} >1$ to $R_{AA}<1$ is observed near $\text{q}_T = 2$ GeV. This value depends on various factors especially on the temperature of the medium. In this analysis, it is found that at lower temperatures, the transition shifts to low $\text{q}_T$ values. It is important to note that when $R_{AA}$ $>$ 1 or =1, especially at low to medium $\text{q}_T$, is complicated to understand since medium modification, initial state effect, and hadronization can sometimes cancel their impacts on $R_{AA}$. One of the possible explanations is that the total number of HQs remains conserved during the evolution, it leads to a shift of HQs at a lower momentum range, causing enhanced $R_{AA}$ at low $\text{q}_T$. 

{
In Fig.~\ref{fig:RAA_C}, at fixed values of temperatures, i.e. $T$= 0.16 GeV (left panel), 0.36 GeV (middle panel) and 0.48 GeV (right panel), we have observed a similar pattern for the suppression of $R_{AA}$. In all three cases, for the highest value of the magnetic field, $eB$ = 0.6 GeV$^2$, we have found the maximum suppression as compared to the vanishing magnetic field ($eB$ = 0.0). On the contrary, in the case of a low magnetic field, i.e. $eB$ =0.01 GeV$^2$, the difference with $eB$ = 0.0 is around $2\sim 3 ~\%$ at $T =0.16$ GeV and lesser at higher temperatures. Interestingly, the MC and IMC effects give small but finite contributions only at the lower temperature, $T=0.16$ GeV. However, these effects are almost absent at higher temperatures considered in the present study, i.e. $T=0.36$ and $T=0.48$ GeV.
}

% Now, for the highest magnetic field, $eB = 0.6$ GeV$^2$ and temperature, $T$ = 0.16 GeV, we have found the maximum suppression as compared to the vanishing magnetic field ($eB$ = 0.0) since the MC and IMC are more prominent at lower temperatures, as shown in Fig. \ref{fig:EL_T16}. However, the suppression at the same $eB$ and $T$= 0.36 GeV, and $T$ = 0.48 GeV is comparatively low. In the case of a low magnetic field, $eB$ =0.01 GeV$^2$ the suppression is around $2\sim 3 ~\%$ at $T$ =0.16 GeV and lesser at higher temperatures.

The results show that it is crucial to include the magnetic field in the analysis to determine the energy loss experienced by HQs as Fig.~\ref{fig:RAA_C} illustrates that the magnetic field substantially affects HQ transport in the QGP medium alongside polarisation processes. Therefore, considering the magnetic field is crucial to ensure theoretical consistency in characterizing HQ transport in the QGP medium.

\section{Summary and future aspects}
\label{el:SaF}
We have studied the dynamics of HQs within the framework of classical Wong's equation. In this study, we have focused on the energy loss due to the medium polarization as the primary factor responsible for the reduction in energy experienced by HQs. Here, the energy dissipation is caused by HQ's soft momentum transfer while traversing through the hot QGP medium in the presence of an intense background magnetic field. Additionally, we have evaluated the influence of different magnetic field strengths on $R_{AA}$ of HQs. Our findings suggest that the magnetic field significantly impacts the energy loss experienced by HQs in the QGP.
Our observations indicate that there is an increase in energy loss with an increase in magnetic field strength. Furthermore, we have noted that the $R_{AA}$ exhibits more suppression as the magnitude of the magnetic field increases.
\\
It is important to note that in the present analysis, we have the fixed values of $eB$ and $T$, which is an assumption to the actual scenario and may not fully capture the dynamic nature of these parameters during the course of heavy ion collisions at RHIC and LHC. It is because, in reality, both $T$ and $eB$ vary with time, especially the magnetic field rapidly decays after the collision and the system cools during the subsequent expansion. In the current analysis, incorporating time-varying $T$ and $eB$ is a complex task, particularly when considering the static scenario that we are investigating. The rapid decay of $eB$ poses a significant challenge, leading to a situation where only a very weak $eB$ persists at later stages. Furthermore, the variation in $eB$ generates an electric field, and when $eB$ becomes weak, a strong electric field emerges, as demonstrated in Ref. \cite{Das:2016cwd}. While these effects are intriguing, they extend beyond the scope of our present analysis. Therefore, we only explore the implications of a wide range of static $eB$ and $T$, as detailed in Refs. \cite{Machado:2013rta, Gubler:2015qok}.
Next, the current analysis only takes into account the energy loss caused by the medium polarization with soft momentum transfer that can be extended with the incorporation of field fluctuations in the medium. In addition, some other future extensions of the current work could include momentum anisotropy as well as viscous effects and the incorporation of a general magnetic field through the gluon propagator for the hot QCD/QGP medium.

\section{Acknowledgements}
 MY Jamal would like to acknowledge the SERB-NPDF (National postdoctoral fellow) File No. PDF/2022/001551. J. Prakash acknowledges the support from DAE-BRNS, India, Grant No. 57/14/02/ 2021-BRNS. I Nilima acknowledge the Women Scientist Scheme A (WoS A) of the Department of Science and Technology (DST) for the funding with grant no. DST/WoS-A/PM-79/2021.  A.B. acknowledges the support from the Alexander von Humboldt Foundation postdoctoral research fellowship in Germany.

\end{document}